\documentclass[manuscript]{aastex63}
\usepackage{amsmath}

\newcommand{\2}{$_2$}

\submitjournal{ApJL}

\shorttitle{}
\shortauthors{}

\begin{document}

\title{Stabilization of dayside surface liquid water via tropopause cold trapping on arid \added{slowly rotating} tidally locked planets}

\correspondingauthor{Feng Ding}
\email{fengding@g.harvard.edu}

\author[0000-0001-7758-4110]{Feng Ding}
\affiliation{School of Engineering and Applied Sciences, Harvard University, Cambridge, MA 02138, USA}

\author[0000-0003-1127-8334]{Robin D. Wordsworth}
\affiliation{School of Engineering and Applied Sciences, Harvard University, Cambridge, MA 02138, USA}
\affiliation{Department of Earth and Planetary Sciences, Harvard University, Cambridge, MA 02138, USA}



\begin{abstract}
Terrestrial-type exoplanets orbiting nearby red dwarf stars (M-dwarfs) are among the best targets for atmospheric characterization and biosignature searches in the near future. Recent evolutionary studies have suggested that terrestrial planets in the habitable zone of M-dwarfs are probably tidally locked and have limited surface water inventories as a result of their host stars' high early luminosities. Several previous climate simulations of such planets have indicated that their remaining water would be transported to the planet's permanent nightside and become trapped as surface ice, leaving the dayside devoid of water. Here we use a three-dimensional general circulation model with a water cycle and accurate radiative transfer scheme to investigate the surface water evolution on \added{slowly rotating} tidally locked terrestrial planets with limited surface water inventories. We show that there is a competition for water trapping between the nightside surface and the substellar tropopause in this type of climate system. Although under some conditions the surface water remains trapped on the nightside as an ice sheet, in other cases liquid water stabilizes in a circular area in the substellar region as a \replaced{shallow sea}{wetland}. Planets with 1 bar N\2 and atmospheric CO\2 levels greater than 0.1 bar retain stable dayside liquid water, even with very small surface water inventories. Our results reveal the diversity of possible climate states on terrestrial-type exoplanets and highlight the importance of surface liquid water detection techniques for future characterization efforts.

\end{abstract}



\section{Introduction} \label{sec:intro}
Terrestrial planets around M-dwarfs are the first potentially habitable exoplanets for which atmospheric characterization will be possible because of their enhanced transit probability, large transit depth and high planet-star contrast ratio \citep{seager2010exoplanet}. However, the climate evolution of this type of planet likely differs from that of the Earth in several important aspects \citep{shields2016habitability}. First, potentially habitable exoplanets around M-dwarfs are probably tidally locked by gravitational tidal torques. In particular, planets with small eccentricities are likely to be trapped in a 1:1 spin-orbit resonance \citep{Kasting1993habitable, barnes2017tidal}. In addition, many of these planets may be deficient in water, because they receive an insolation above the runaway greenhouse threshold during their host stars' pre-main-sequence phase, leading to extensive water loss \citep{ramirez2014premain, luger2015waterloss, tian2015waterloss}. Studies that assumed an Earth-like atmospheric composition have shown that for such planets, water vapor would be transported by the atmospheric circulation to the cold permanent nightside and trapped as surface ice \citep{heath1999habitability, joshi2003synchronously, menou2013watertrap, yang2014watertrap}. However, the water cycle on these planets over a wider range of atmospheric parameters has not yet been investigated in detail.

\section{Moist general circulation model} \label{sec:methods}
Here we investigate the water trapping on \added{slowly rotating} tidally locked planets with limited surface water inventories (referred to as ``arid planets'' in comparison with the ``aqua-planet'' simulations widely used in previous studies) over a range of atmospheric compositions. We developed a three-dimensional general circulation model (GCM) with a self-consistent water cycle based on our dry GCM that uses a line-by-line approach to describe the radiative transfer for simulating diverse planetary atmospheres \citep{ding2019fmspcm}. The physical schemes for moist processes are very similar to those used in \citet{merlis2010tidally}, including a moist convection scheme, a large-scale condensation scheme and a planetary boundary scheme.  Our moist GCM can reproduce the climatology on tidally locked Earth-like aqua-planets  in \citet{merlis2010tidally} when using the same gray-gas radiative transfer calculation.

To simulate the climate of tidally locked terrestrial planets around M-dwarfs, we use AD Leo’s stellar spectrum. The planet has the same radius and surface gravity as Earth's and an orbital period of 35 Earth days. Both the eccentricity and obliquity are zero. The incoming stellar radiation above the substellar point is 1200 W m$^{-2}$ \deleted{ and the surface albedo is 0.25}. \added{The surface is flat and has a albedo of 0.25.} 2000 spectral points and four quadrature points (two upwelling and two downwelling) are used for both shortwave and longwave radiative calculations.
The atmosphere is made of N\2, CO\2 and H\2O. The column-integrated mass of N\2 is equivalent to the mass of a 1 bar N\2 atmosphere if the atmosphere was made of N\2 alone. We performed simulations with various CO\2 levels and found the surface water starts to be converged to the substellar region when the CO\2 mass mixing ratio is 0.1. So we chose two CO\2 levels to present our results and discuss the two climate regimes: (1) one has a present-day Earth-like CO\2 level with the CO\2 volume mixing ratio of 400 ppmv (referred to as the ``low CO\2 run''); (2) the other has a relatively higher CO\2 level and the column-integrated mass of CO\2 is equivalent to the mass of a 0.1 bar CO\2 atmosphere if the atmosphere was made of CO\2 alone (referred to as the ``high CO\2 run''). Thus, for the latter case, CO\2 also contributes substantially to the total mass of the atmosphere. The total surface pressure is 1.1 bar with a CO\2 volume mixing ratio of 0.063 relative to N\2. In both simulations, the water vapor mixing ratio is controlled by the atmospheric circulation and surface water distribution.

To simulate the surface water evolution, we implemented a bucket water model in which the water depth of the bucket varies with the local precipitation and evaporation (see Appendix \ref{app:bucket} for details). The initial condition of the surface water inventory is 1 m of ice uniformly distributed on the nightside surface. The GCM simulations are run until reaching the statistical equilibrium state ($\sim$3000 days) when the absorbed stellar radiation of the climate system is balanced by outgoing longwave radiation and the surface water distribution stops changing with time. Averaged results over the last 600 days are presented.


\section{Equilibrium distribution of surface water inventory and water trapping competition} \label{sec:equilibrium}

\begin{figure}[ht]
  \centering
  \includegraphics[width=\columnwidth]{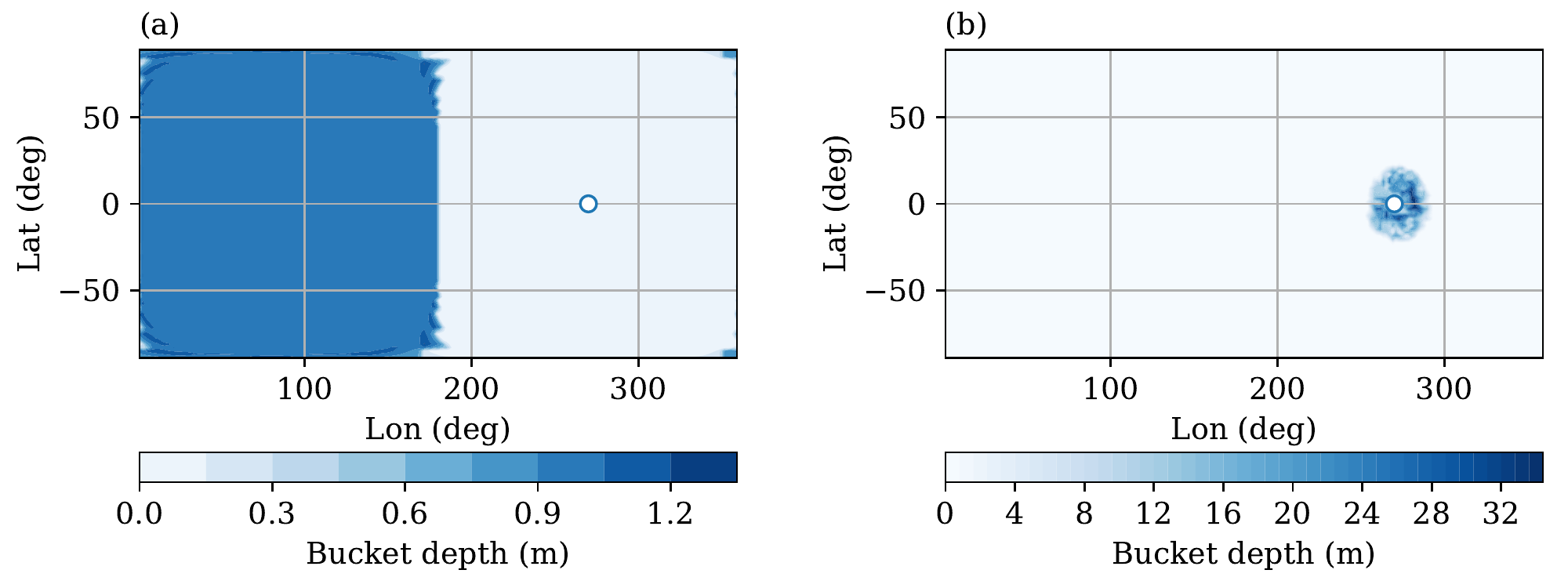}
  \caption{Long-term mean surface water depth for the low CO\2 run with volume mixing ratio of 400 ppmv (a) and for the high CO\2 run with volume mixing ratio of 0.063 (b). The white dot is the substellar point.}\label{fig:depth}
\end{figure}

Fig.~\ref{fig:depth} shows the distributions of the surface water depth in the equilibrium state of the two simulations, after the surface water evolution algorithm had converged. For the low CO\2 run, the surface water is stably trapped on the nightside as an ice sheet (Fig.~\ref{fig:depth}a). But for the high CO\2 run, the initial nightside surface ice migrates towards the substellar area and eventually forms a \replaced{shallow sea}{wetland} surrounded by a hyper-arid desert on the dayside. This substellar ``oasis'' occupies a circular area roughly 20$^\circ$ around the substellar point (Fig.~\ref{fig:depth}b). The formation of this oasis results from the moist climate dynamics, specifically, the cold trapping effect of water vapor near the substellar tropopause and the associated precipitation. In Fig.~\ref{fig:depth}b, the center of the oasis is slightly shifted eastward relative to the substellar point, because the upwelling motion west of the substellar point is suppressed by planetary equatorial waves \citep{yang2013innerhz}.

\begin{figure}[ht]
  \centering
  \includegraphics[width=\columnwidth]{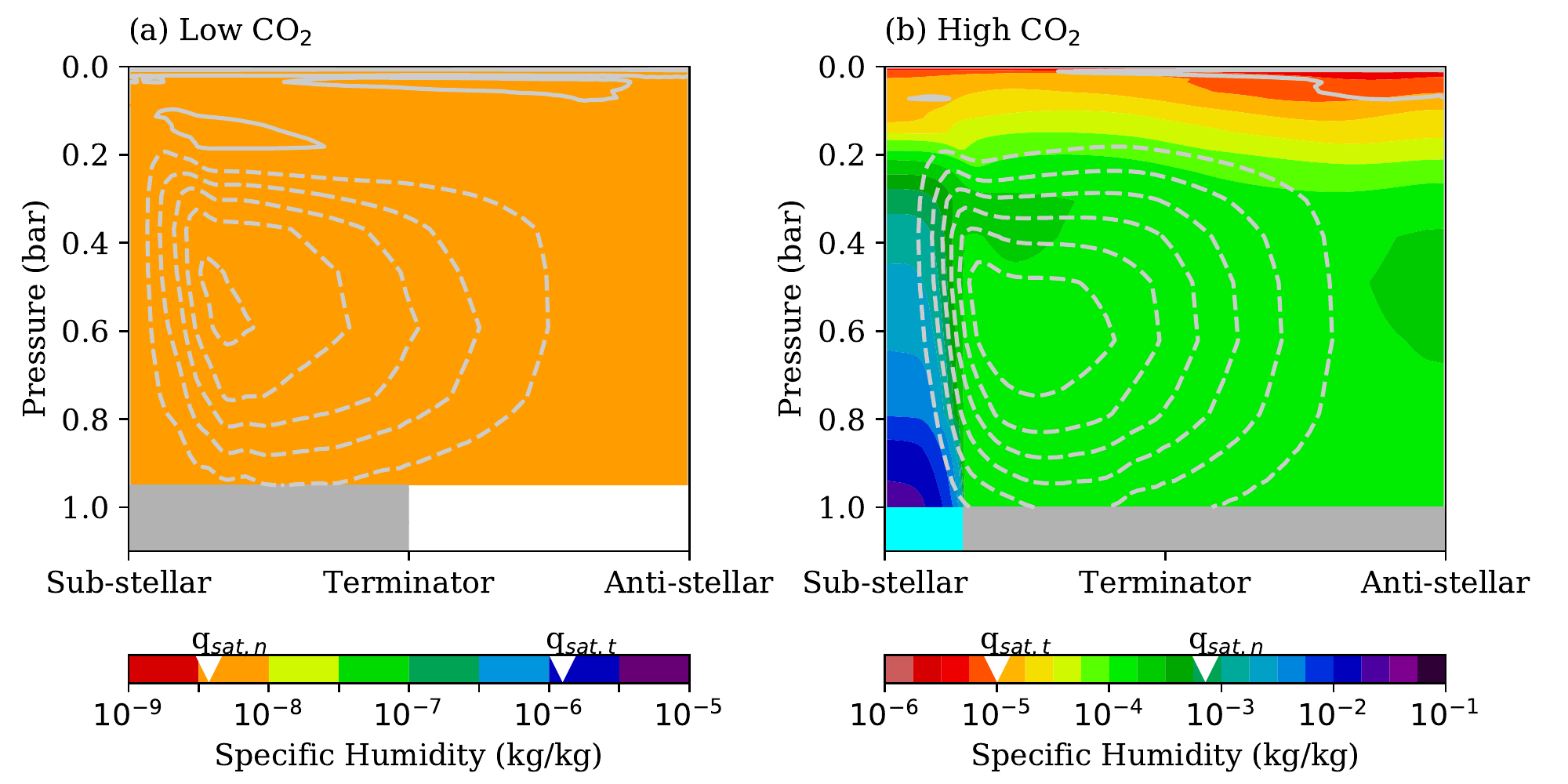}
  \caption{Zonal mean distribution of the specific humidity in the tidally locked coordinate for the low CO\2 run (a) and high CO\2 run (b). q$_{sat,n}$ and q$_{sat,t}$ above the color bar are the saturation specific humidity calculated by the conditions on the nightside surface and at the substellar tropopause, respectively. The dotted gray, white and light blue area at the bottom in the two panels mark the distribution of the dry surface, nightside surface ice in (a) and the substellar open water in (b), respectively. The gray dashed contours in the two panels mark the tropospheric overturning circulation where air rises in the substellar region.}\label{fig:q}
\end{figure}

By comparing the two simulations with different CO\2 levels, we can elucidate how the cold trapping competition between the nightside surface and the substellar tropopause determines the equilibrium distribution of the surface water inventory on a tidally locked arid planet. To better illustrate this competition, we plot the specific humidity (q, in kg/kg) distribution in the tidally locked coordinate \footnote{\url{https://github.com/ddbkoll/tidally-locked-coordinates}} \citep{koll2015phasecurve, wordsworth2015heat}  in Fig.~\ref{fig:q} and we use the saturation specific humidity (q$_{sat}$) calculated by the conditions on the nightside surface (q$_{sat,n}$) and at the substellar tropopause (q$_{sat,t}$) to evaluate the strength of water vapor cold trapping. For the low CO\2 run, q$_{sat,n}$ $ \approx 3.74\times10^{-9}$, q$_{sat,t}$ $\approx 1.25\times10^{-6}$ and q$_{sat,n}$ $\ll$ q$_{sat,t}$. Therefore, the atmospheric water vapor is constrained by the nightside surface ice sheet and well mixed in the atmosphere without any condensation (Fig.~\ref{fig:q}a). But for the high CO\2 run, q$_{sat,n}$ $\approx 7.2\times10^{-4}$, q$_{sat,t}$ $\approx 1\times10^{-5}$. Precipitation forms within the upwelling branch of the overturning circulation, giving rise to an open water area with the same size as the area where air rises. In Fig.~\ref{fig:q}b, the atmospheric profile follows the moist adiabat above the open water, but once the air parcel leaves the upwelling branch of the overturning cell condensation ceases, and the water vapor concentration becomes well mixed again. In fact, most water vapor on the nightside is not last saturated exactly at the substellar tropopause but in a region around the tropopause, shown as the green shading area in Fig.~\ref{fig:q}b. Hence, the nightside atmospheric specific humidity is slightly higher than q$_{sat,t}$ but still well below q$_{sat,n}$. As a result, any nightside surface water inventory slowly sublimates into the atmosphere, explaining the migration of surface water from nightside to the substellar area. This occurs despite the fact that the nightside thermal stratification is very stable due to the strong near-surface temperature inversion. 

\begin{figure}[ht]
  \centering
  \includegraphics[width=0.5\columnwidth]{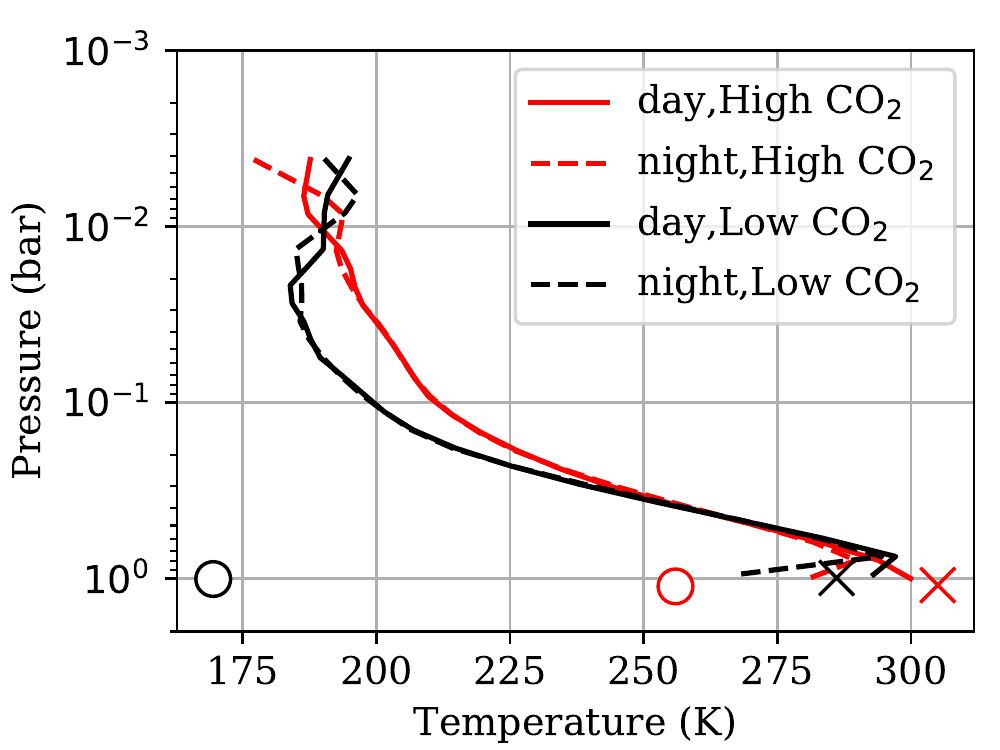}
  \caption{Hemispheric-averaged dayside (solid) and nightside (dashed) vertical temperature profiles in the low CO\2 (black) and the high CO\2 (red) simulations. The hemispheric-averaged day and nightside surface temperatures are marked by crosses and circles, respectively.}\label{fig:t}
\end{figure}

In the high CO\2 run, the substellar tropopause wins in the competition of water vapor cold trapping mainly because of the radiative effect of CO\2. The ratio of the two saturation specific humidities at the two cold traps can be written as

\begin{equation} \label{eq:coldtrap}
\frac{q_{sat,t}}{q_{sat,n}} \approx \frac{e_{sat}(T_{trop})}{e_{sat}(T_{sn})} \frac{p_{sn}}{p_{trop}}
\end{equation}

where $e_{sat}$ is the saturation vapor pressure of water, $T_{trop}$ and $p_{trop}$ are the air temperature and pressure at the substellar tropopause,  and $T_{sn}$ and $p_{sn}$ are the surface temperature and pressure  on the nightside. Both $p_{sn}$ and $e_{sat}(T_{trop})$ increase a little in the high CO\2 run due to the mass contribution and near-infrared absorption of CO\2. But the major impact of the high CO\2 level is warming of the nightside surface temperature from 170 K under the present-day CO\2 level to 255~K when the CO\2 volume mixing ratio is 0.063 (Fig.~\ref{fig:t}). The saturation surface specific humidity is therefore raised by nearly five orders of magnitude, after which point it exceeds the value around the substellar tropopause (Fig.~\ref{fig:q}). The enhanced warming on the nightside surface is attributed to the enhanced infrared emissivity of the atmosphere, not only from the high CO\2 level, but also from the higher water vapor concentration (Fig.~\ref{fig:q}). This warming mechanism has been discussed previously in the context of completely dry atmospheres \citep{wordsworth2015heat, koll2016heat, ding2019fmspcm}. Another factor that can potentially affect the nightside surface temperature is turbulent mixing in the planetary boundary layer. However, in our high CO\2 run, the sensible heat flux from the atmosphere to the nightside surface is less than 5 W m$^{-2}$, while the infrared radiative flux reaching the nightside surface is $\sim$ 230 W m$^{-2}$. Hence, the nightside surface is mainly in radiative equilibrium and the turbulent heat mixing plays a minor role here.

\section{CO\2 budget and stability of the two equilibrium climate states} \label{sec:co2}
So far, we have shown the two equilibrium climate states under different but fixed CO\2 levels. However, the CO\2 partial pressure on Earth is regulated by the silicate-weathering feedback \citep{walker1981co2}, and for many terrestrial exoplanets a similar feedback may be important. Although it is challenging to evaluate the outgassing and weathering of CO\2 on exoplanets in general, we can gain insight into the CO\2 budget of the two equilibrium climate states qualitatively by using a simple equation for the partial pressure of CO\2. In this simple CO\2 budget model, the surface partial pressure of CO\2 is determined by outgassing and weathering, and the CO\2 weathering rate is primarily determined by the CO\2 surface partial pressure and the surface area fraction of liquid water. 
\begin{equation} \label{eq:co2}
\frac{dP_{co2}}{dt} = \begin{cases} 
		\Phi,  & P_{co2}<0.07\ \mathrm{bar}\  (\mathrm{nightside\ cold\ trap}) \\
		\Phi-W_0e^{k[T_s(P_{co2})-T_0]} A_{frac}. &  P_{co2} \ge 0.07\ \mathrm{bar}\  (\mathrm{substellar\ tropopause\ cold\ trap}) \\
	\end{cases} \\
\end{equation}
where $P_{co2}$ is the surface CO\2 partial pressure, $\Phi$ is the CO\2 outgassing rate, $W_0$ = 70 bar Gyr$^{-1}$, $T_0$ = 288 K are the CO\2 weathering rate and surface temperature in the reference state, k = 0.1 K$^{-1}$ is the weathering-temperature rate constant \citep{abbot2016co2}, and Ts, Afrac = 0.03 are the substellar surface temperature and fraction of the substellar liquid water area, respectively. In our CO\2 budget model, the weathering rate depends on the surface CO\2 partial pressure indirectly through the substellar surface temperature. The direct power law dependence on the surface CO\2 partial pressure is ignored here, because recent work based on laboratory experiments that have quantified the dissolution rate of a variety of silicate minerals has indicated that this dependence is weak \citep{graham2019pressure}.

In our high CO\2 run, the surface partial pressure of CO\2 is 1.1$\times$0.063=0.07 bar. So we use a CO\2 partial pressure of 0.07 bar as the threshold above which surface water migrates from the nightside to the substellar area and then weathering occurs. The timescale of the surface water migration is short compared to the weathering timescale, so we assume that the surface water is always in an equilibrium state in this model. When the surface CO\2 partial pressure is 0.07 bar, a critical CO\2 outgassing rate of $\Phi_c$=52 bar Gyr$^{-1}$ is found in this CO\2 budget model by which two climate regimes can be discriminated. If the CO\2 outgassing rate is less than $\Phi_c$, the climate on tidally locked arid planets has no stable equilibrium state, and the surface water will oscillate between the nightside ice and substellar water states, in a pattern that resembles the limit cycles predicted to occur for some planets near the outer edge of habitable zone \citep{haqq2016limitcycle}. For CO\2 outgassing rate larger than $\Phi_c$, the substellar water solution becomes a stable climate state because the weathering rate can always balance the outgassing rate through the CO\2 partial pressure dependence. Interestingly, the critical value of the CO\2 outgassing rate found in our simple formula is close to Earth's CO\2 outgassing rate. In reality, weathering is affected by many complicated processes, including uplift rates, the planetary tectonic regime, surface maficity and other factors \citep{macdonald2019arc, oneill2007superearth, valencia2007superearth}, and hence can be expected to vary over a wide range.

\section{Three climate regimes under increased insolation} \label{sec:evo}

\begin{figure}[ht]
  \centering
  \includegraphics[width=\columnwidth]{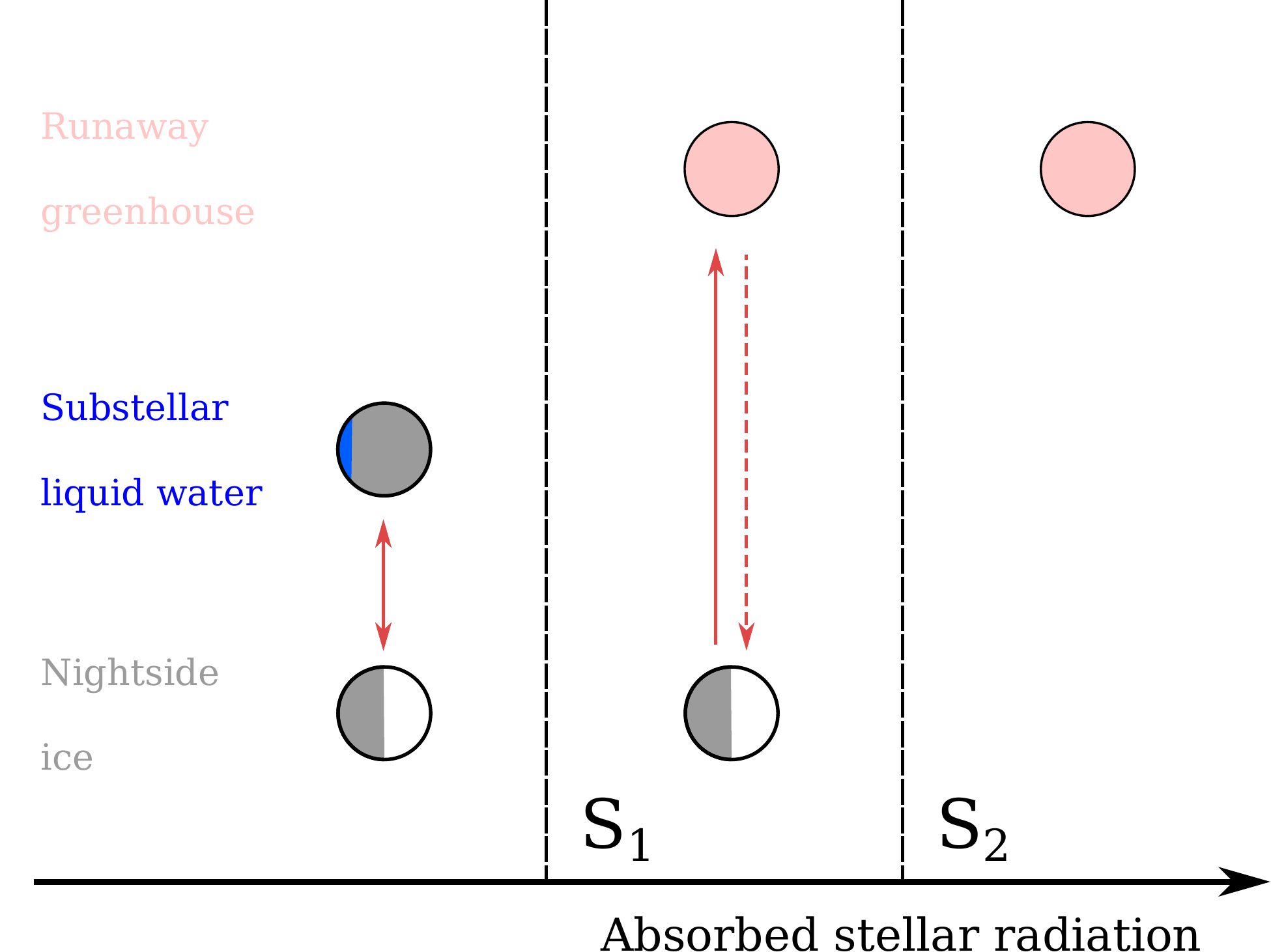}
  \caption{Schematic of the long-term climate evolution and possible climate states on a tidally locked arid planet receiving various insolations. The processes that lead to climate transitions between various states are marked by red arrows, e.g., the CO\2 cycle (solid) and impact erosion (dashed).  S$_1$ is the critical absorbed stellar flux when the substellar liquid water state enters the runaway greenhouse, which is $\sim$\replaced{1100}{286} W m$^{-2}$ \citep{yang2013innerhz}. S$_2$ is the critical absorbed stellar flux when the nightside ice state enters the runaway greenhouse, $\sim$ \replaced{1550}{388} W m$^{-2}$ in our simulation.}\label{fig:evo}
\end{figure}

The two simulations we have discussed with varying CO\2 levels were performed under an Earth-like insolation level. Because planets closer to their host stars are more observationally favorable, we also investigated how the substellar surface water and nightside surface ice state evolve when the arid planet receives increased insolation. Similar to aqua-planets, arid planets with stable substellar surface water should enter the runaway greenhouse state when the absorbed stellar radiation of the planet exceeds the upper limit of the outgoing longwave radiation that the planet can emit. We did not attempt to simulate this climate transition in our current model because water vapor is a non-dilute component when approaching the runaway greenhouse state for which special treatment is required \citep{ding2016condensible, pierrehumbert2016nondilute}. However, based on previous simulations of the runaway greenhouse on tidally locked aqua-planets \citep{yang2013innerhz}, the critical absorbed stellar flux for this transition on Earth-like tidally locked planets can be estimated as S$_1 \sim$\replaced{1100}{286} W m$^{-2}$.  The nightside surface temperature also increases with the insolation. In our simulations with enhanced insolation and stable nightside ice, the ice begins to irreversibly sublimate into the atmosphere when the absorbed stellar radiation exceeds S$_2 \sim$  \replaced{1550}{388} W m$^{-2}$.

To summarize, three climate regimes on a tidally locked arid planet can be discriminated by these two critical fluxes S$_1$ and S$_2$ as illustrated in Fig.\ref{fig:evo}. When the absorbed stellar radiation of the climate system S$_{abs} <$ S$_1$, the surface water can either distribute on the nightside as ice sheet or in the substellar area. These two states can transfer from one to the other via changes in the carbonate-silicate cycle, as discussed in Section~\ref{sec:co2}. When S$_1 <$ S$_{abs} <$ S$_2$, surface water can only exist as an ice sheet on the nightside. A runaway greenhouse will occur if the cold trap on the nightside surface is weaker, for example, due to nightside surface warming by CO\2. This is similar to the bi-stable moist climate states discussed for close-in arid exoplanets \citep{leconte2013bistable}. Finally, when S$_{abs} >$ S$_2$, all water is present in the atmosphere as vapor, and no surface water is possible. On long timescales this atmospheric water vapor would be vulnerable to photodissociation and hydrogen loss to space \citep{kasting1988runaway}.

\section{Conclusion and discussions} \label{sec:discussions}
Our results demonstrate that even for tidally locked arid exoplanets with low water inventories, climate solutions exist where dayside surface liquid water is stable. This implies habitable conditions on planets around M-stars may be more common than previously considered. Future research efforts will need to focus on the effects of other volatile cycles and water-rock interactions on the climates of such planets, as well as on developing robust observational techniques to constrain the abundance of surface liquid water remotely \citep{robinson2010glint, loftus2019sulfate}. 

\added{Other than the CO\2 cycle discussed in Section~\ref{sec:co2}, the stability of the dayside surface water can be potentially affected by some other factors not included in our idealized GCM simulations. First, for higher surface water inventories well beyond 1m nightside equivalent, surface water would flow away from the substellar region and refreeze on the nightside. The critical surface water inventory under which surface water is stably trapped on the dayside should be investigated in the future with a more complex land model that incorporates both surface runoff and evaporation.
Second, the stabilization of the dayside surface water also relies on the dayside orography. In general, moderately lower elevations (e.g., crater depressions) in the substellar region should help to stabilize  surface water there. Third, fast planetary rotation is likely to alter the proposed mechanism that converges precipitation in the substellar area. For tidally locked terrestrial planets with rotation period shorter than 20 days,  the  thermally direct circulation is greatly disturbed by planetary waves \citep{haqqmisra2018regimes}, and the atmospheric circulation may transport water vapor to the nightside by other pathways without passing through the tropopause cold trap. At last, }
same as in \citet{merlis2010tidally}, our GCM is cloud-free to allow a focus on the key physical processes (e.g., the radiative impact of greenhouse gases) without invoking cloud parameterizations. Cloud effects involve various atmospheric processes over a wide range of spatial scales and always need to be parameterized in large-scale climate models, which remain a major problem even for modern climate change simulations \citep{stephens2005cloud}. Despite the challenges associated with cloud modeling in GCMs, via basic reasoning we can establish that cloud radiative effects would have limited impact on the water trapping competition on arid tidally locked planets. When the surface water is trapped on the nightside as ice, clouds would be scarce in the atmosphere and could not contribute to the migration of surface water towards the substellar region. In the climate state with substellar surface liquid water, the troposphere above the surface water region would be covered by deep convective clouds that are highly reflective. The upper troposphere would be covered by high cirrus clouds that could extend to the nightside
, as seen in some aqua-planet climate simulations \citep{yang2013innerhz}. These cirrus clouds would warm the nightside surface, making the dayside tropopause cold trapping more robust. The substellar surface water only occupies a very small fraction of the planet’s surface area (Figs~\ref{fig:depth} and \ref{fig:q}), where only 11\% of incoming radiation is received in total. So even if the deep convective clouds increased the local albedo to 70\% above the dayside liquid water region, the total planetary albedo would only be raised by 6\%, which would have a limited effect on the global mean surface temperature. Even if inclusion of clouds in the simulations somehow induced a more significant cooling effect, it would only mean that a somewhat higher CO\2 mixing ratio would be required to drive the surface water from the nightside to dayside.

In situations where the substellar tropopause cold trap dominates the hydrological cycle, the climate regime we find is a natural extension of Earth's tropical climate in which the Hadley circulation transports surface moisture towards the equator and then precipitation forms within the upwelling branch of the overturning cell. The stable substellar water region resembles Earth's tropical rain belt (Intertropical Convergence Zone, ITCZ) and the surrounding dry land resembles Earth's subtropical deserts. Similar comparisons to our results can be made with Titan, where a hydrological cycle operates with methane as the condensible component. In Titan's case, the seasonal cycle is much stronger due to the small thermal inertia of the lower boundary and strong seasonal migration of the ICTZ gives rise to Titan's dry tropics and polar lakes \citep{mitchell2016titan}.

\acknowledgments
We thank the referee for thoughtful comments that improved the manuscript. 
The GCM simulations in this paper were carried out on the FASRC Cannon cluster supported by the FAS Division of Science, Research Computing Group at Harvard University. F.D. and R.W. were supported by NASA grants NNX16AR86G and 80NSSC18K0829.



\appendix

\section{Bucket water model iteration} \label{app:bucket}
\begin{figure}[ht]
  \centering
  \includegraphics[width=\columnwidth]{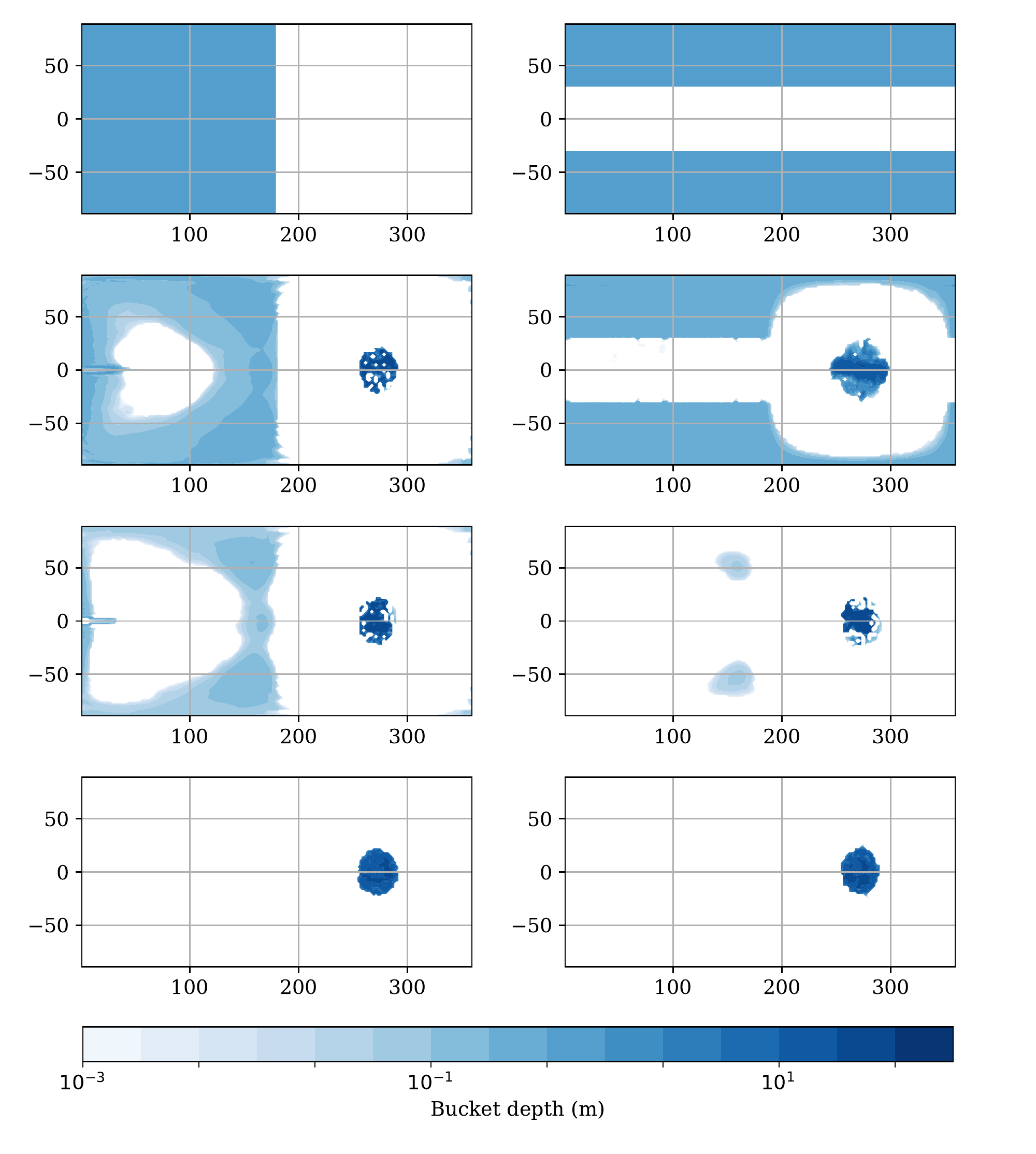}
  \caption{Evolution of the surface water inventory in the high CO\2 run. The total model integration time is 3000 days. The surface water iteration is performed every 200 days with a 10-year timestep. Left and right columns are the snapshots for the two cases with initial surface water on the nightside surface and in the extratropical region, respectively. }\label{fig:bp_time}
\end{figure}

\added{In the bucket model, the rate of change of water depth ($\dot{h}$) depends on the local precipitation ($P$) and evaporation ($E$), $\dot{h} = P-E$. If the total amount of water cannot provide the  evaporation flux computed by the bulk aerodynamic formula in the planetary boundary layer scheme, all water in the bucket will be evaporated instead.}
Because the evolution of the surface water inventory is a much slower process than the atmospheric dynamics, we implemented an iteration scheme for the surface water evolution. The time-mean tendency of water depth in the bucket model is evaluated every 200 days and is used to update the surface water distribution by multiplication with a 10-year timestep, similar to the algorithm described in \citet{wordsworth2013mars} to explore surface ice evolution on early Mars. To validate the surface water iteration scheme, we performed a simulation for the high CO\2 run with different initial surface water distributions. When the surface water is initially distributed in the extratropical region (upper right panel in Fig.~\ref{fig:bp_time}), the climate evolves towards the same equilibrium climate state and the same substellar “oasis” forms.

%
%

%

\bibliography{fmspcm}

\begin{thebibliography}{}
\expandafter\ifx\csname natexlab\endcsname\relax\def\natexlab#1{#1}\fi
\providecommand{\url}[1]{\href{#1}{#1}}
\providecommand{\dodoi}[1]{doi:~\href{http://doi.org/#1}{\nolinkurl{#1}}}
\providecommand{\doeprint}[1]{\href{http://ascl.net/#1}{\nolinkurl{http://ascl.net/#1}}}
\providecommand{\doarXiv}[1]{\href{https://arxiv.org/abs/#1}{\nolinkurl{https://arxiv.org/abs/#1}}}

\bibitem[{{Abbot}(2016)}]{abbot2016co2}
{Abbot}, D.~S. 2016, \apj, 827, 117, \dodoi{10.3847/0004-637X/827/2/117}

\bibitem[{{Barnes}(2017)}]{barnes2017tidal}
{Barnes}, R. 2017, Celestial Mechanics and Dynamical Astronomy, 129, 509,
  \dodoi{10.1007/s10569-017-9783-7}

\bibitem[{{Ding} \& {Pierrehumbert}(2016)}]{ding2016condensible}
{Ding}, F., \& {Pierrehumbert}, R.~T. 2016, \apj, 822, 24,
  \dodoi{10.3847/0004-637X/822/1/24}

\bibitem[{{Ding} \& {Wordsworth}(2019)}]{ding2019fmspcm}
{Ding}, F., \& {Wordsworth}, R.~D. 2019, \apj, 878, 117,
  \dodoi{10.3847/1538-4357/ab204f}

\bibitem[{{Graham} \& {Pierrehumbert}(2019)}]{graham2019pressure}
{Graham}, R.~J., \& {Pierrehumbert}, R. 2019, in AGU Fall Meeting Abstracts,
  Vol. 2019, P21A--02

\bibitem[{{Haqq-Misra} {et~al.}(2016){Haqq-Misra}, {Kopparapu}, {Batalha},
  {Harman}, \& {Kasting}}]{haqq2016limitcycle}
{Haqq-Misra}, J., {Kopparapu}, R.~K., {Batalha}, N.~E., {Harman}, C.~E., \&
  {Kasting}, J.~F. 2016, \apj, 827, 120, \dodoi{10.3847/0004-637X/827/2/120}

\bibitem[{{Haqq-Misra} {et~al.}(2018){Haqq-Misra}, {Wolf}, {Joshi}, {Zhang}, \&
  {Kopparapu}}]{haqqmisra2018regimes}
{Haqq-Misra}, J., {Wolf}, E.~T., {Joshi}, M., {Zhang}, X., \& {Kopparapu},
  R.~K. 2018, \apj, 852, 67, \dodoi{10.3847/1538-4357/aa9f1f}

\bibitem[{{Heath} {et~al.}(1999){Heath}, {Doyle}, {Joshi}, \&
  {Haberle}}]{heath1999habitability}
{Heath}, M.~J., {Doyle}, L.~R., {Joshi}, M.~M., \& {Haberle}, R.~M. 1999,
  Origins of Life and Evolution of the Biosphere, 29, 405

\bibitem[{{Joshi}(2003)}]{joshi2003synchronously}
{Joshi}, M. 2003, Astrobiology, 3, 415, \dodoi{10.1089/153110703769016488}

\bibitem[{{Kasting}(1988)}]{kasting1988runaway}
{Kasting}, J.~F. 1988, \icarus, 74, 472, \dodoi{10.1016/0019-1035(88)90116-9}

\bibitem[{{Kasting} {et~al.}(1993){Kasting}, {Whitmire}, \&
  {Reynolds}}]{Kasting1993habitable}
{Kasting}, J.~F., {Whitmire}, D.~P., \& {Reynolds}, R.~T. 1993, \icarus, 101,
  108, \dodoi{10.1006/icar.1993.1010}

\bibitem[{{Koll} \& {Abbot}(2015)}]{koll2015phasecurve}
{Koll}, D.~D.~B., \& {Abbot}, D.~S. 2015, \apj, 802, 21,
  \dodoi{10.1088/0004-637X/802/1/21}

\bibitem[{{Koll} \& {Abbot}(2016)}]{koll2016heat}
---. 2016, \apj, 825, 99, \dodoi{10.3847/0004-637X/825/2/99}

\bibitem[{{Leconte} {et~al.}(2013){Leconte}, {Forget}, {Charnay}, {Wordsworth},
  {Selsis}, {Millour}, \& {Spiga}}]{leconte2013bistable}
{Leconte}, J., {Forget}, F., {Charnay}, B., {et~al.} 2013, \aap, 554, A69,
  \dodoi{10.1051/0004-6361/201321042}

\bibitem[{{Loftus} {et~al.}(2019){Loftus}, {Wordsworth}, \&
  {Morley}}]{loftus2019sulfate}
{Loftus}, K., {Wordsworth}, R.~D., \& {Morley}, C.~V. 2019, \apj, 887, 231,
  \dodoi{10.3847/1538-4357/ab58cc}

\bibitem[{{Luger} \& {Barnes}(2015)}]{luger2015waterloss}
{Luger}, R., \& {Barnes}, R. 2015, Astrobiology, 15, 119,
  \dodoi{10.1089/ast.2014.1231}

\bibitem[{{Macdonald} {et~al.}(2019){Macdonald}, {Swanson-Hysell}, {Park},
  {Lisiecki}, \& {Jagoutz}}]{macdonald2019arc}
{Macdonald}, F.~A., {Swanson-Hysell}, N.~L., {Park}, Y., {Lisiecki}, L., \&
  {Jagoutz}, O. 2019, Science, 364, 181, \dodoi{10.1126/science.aav5300}

\bibitem[{{Menou}(2013)}]{menou2013watertrap}
{Menou}, K. 2013, \apj, 774, 51, \dodoi{10.1088/0004-637X/774/1/51}

\bibitem[{{Merlis} \& {Schneider}(2010)}]{merlis2010tidally}
{Merlis}, T.~M., \& {Schneider}, T. 2010, Journal of Advances in Modeling Earth
  Systems, 2, 13, \dodoi{10.3894/JAMES.2010.2.13}

\bibitem[{{Mitchell} \& {Lora}(2016)}]{mitchell2016titan}
{Mitchell}, J.~L., \& {Lora}, J.~M. 2016, Annual Review of Earth and Planetary
  Sciences, 44, 353, \dodoi{10.1146/annurev-earth-060115-012428}

\bibitem[{{O'Neill} \& {Lenardic}(2007)}]{oneill2007superearth}
{O'Neill}, C., \& {Lenardic}, A. 2007, \grl, 34, L19204,
  \dodoi{10.1029/2007GL030598}

\bibitem[{{Pierrehumbert} \& {Ding}(2016)}]{pierrehumbert2016nondilute}
{Pierrehumbert}, R.~T., \& {Ding}, F. 2016, Proceedings of the Royal Society of
  London Series A, 472, 20160107, \dodoi{10.1098/rspa.2016.0107}

\bibitem[{{Ramirez} \& {Kaltenegger}(2014)}]{ramirez2014premain}
{Ramirez}, R.~M., \& {Kaltenegger}, L. 2014, \apjl, 797, L25,
  \dodoi{10.1088/2041-8205/797/2/L25}

\bibitem[{{Robinson} {et~al.}(2010){Robinson}, {Meadows}, \&
  {Crisp}}]{robinson2010glint}
{Robinson}, T.~D., {Meadows}, V.~S., \& {Crisp}, D. 2010, \apjl, 721, L67,
  \dodoi{10.1088/2041-8205/721/1/L67}

\bibitem[{{Seager} \& {Deming}(2010)}]{seager2010exoplanet}
{Seager}, S., \& {Deming}, D. 2010, \araa, 48, 631,
  \dodoi{10.1146/annurev-astro-081309-130837}

\bibitem[{{Shields} {et~al.}(2016){Shields}, {Ballard}, \&
  {Johnson}}]{shields2016habitability}
{Shields}, A.~L., {Ballard}, S., \& {Johnson}, J.~A. 2016, \physrep, 663, 1,
  \dodoi{10.1016/j.physrep.2016.10.003}

\bibitem[{{Stephens}(2005)}]{stephens2005cloud}
{Stephens}, G.~L. 2005, Journal of Climate, 18, 237,
  \dodoi{10.1175/JCLI-3243.1}

\bibitem[{{Tian} \& {Ida}(2015)}]{tian2015waterloss}
{Tian}, F., \& {Ida}, S. 2015, Nature Geoscience, 8, 177,
  \dodoi{10.1038/ngeo2372}

\bibitem[{{Valencia} {et~al.}(2007){Valencia}, {O'Connell}, \&
  {Sasselov}}]{valencia2007superearth}
{Valencia}, D., {O'Connell}, R.~J., \& {Sasselov}, D.~D. 2007, \apjl, 670, L45,
  \dodoi{10.1086/524012}

\bibitem[{{Walker} {et~al.}(1981){Walker}, {Hays}, \&
  {Kasting}}]{walker1981co2}
{Walker}, J.~C.~G., {Hays}, P.~B., \& {Kasting}, J.~F. 1981, \jgr, 86, 9776,
  \dodoi{10.1029/JC086iC10p09776}

\bibitem[{{Wordsworth}(2015)}]{wordsworth2015heat}
{Wordsworth}, R. 2015, \apj, 806, 180, \dodoi{10.1088/0004-637X/806/2/180}

\bibitem[{{Wordsworth} {et~al.}(2013){Wordsworth}, {Forget}, {Millour}, {Head},
  {Madeleine}, \& {Charnay}}]{wordsworth2013mars}
{Wordsworth}, R., {Forget}, F., {Millour}, E., {et~al.} 2013, \icarus, 222, 1,
  \dodoi{10.1016/j.icarus.2012.09.036}

\bibitem[{{Yang} {et~al.}(2013){Yang}, {Cowan}, \& {Abbot}}]{yang2013innerhz}
{Yang}, J., {Cowan}, N.~B., \& {Abbot}, D.~S. 2013, \apj, 771, L45,
  \dodoi{10.1088/2041-8205/771/2/L45}

\bibitem[{{Yang} {et~al.}(2014){Yang}, {Liu}, {Hu}, \&
  {Abbot}}]{yang2014watertrap}
{Yang}, J., {Liu}, Y., {Hu}, Y., \& {Abbot}, D.~S. 2014, \apjl, 796, L22,
  \dodoi{10.1088/2041-8205/796/2/L22}

\end{thebibliography}
\bibliographystyle{aasjournal}


\listofchanges

\end{document}